\documentclass[11pt]{article}
\usepackage{amssymb,amsfonts,amsmath}
\usepackage[pdftex]{graphicx}
\DeclareGraphicsExtensions{.pdf}  \textheight24.5cm
\begin{document}

\renewcommand{\thefootnote}{\fnsymbol{footnote}}

\centerline{\Large \bf On the use of the incompressibility condition in the} 
\vspace*{1.5mm}
\centerline{\Large \bf Euler and Navier--Stokes equations} 
\vspace*{3mm} \centerline{Peter Stubbe\footnote[1]{retired from Max--Planck--Institut f\"ur Sonnensystemforschung, 37077 G\"ottingen, Germany. Contact by e-mail: peter-stubbe@t-online.de }} 
 
\renewcommand{\thefootnote}{\arabic{footnote}}

\vspace*{.5cm}{\small \begin{quote}  The Euler and Navier--Stokes equations both belong to a closed system of three transport equations, describing the particle number density $N$, the macroscopic velocity {\bf v} and the temperature $T$. These sytems are complete, leaving no room for any additional equation. Nonetheless, it is common practice in parts of the literature to replace the thermal equation by the incompressibility condition $\nabla \cdot {\bf v} = 0$, motivated by the wish to obtain simpler equations. It is shown that this procedure is physically inconsistent in several ways, with the consequence that incompressibility is not a property that can be enforced by an external condition. Incompressible behaviour, if existing, will have to follow self--consistently from the full set of transport equations. \end{quote} }

\vspace*{.5cm}{\large \bf 1. Introduction}

A substance is incompressible if its volume cannot be reduced by exerting a pressure. Obviously, a gas is not incompressible, and yet one finds an abundance of publications in hydrodanamics treating a gas as if it were incompressible. The reasoning behind this is that there may exist processes that can be theoretically described in an acceptable way by treating the gas as if it were incompressible. The task in this case is twofold, firstly to use the incompressibility condition for simplifying the existing hydrodynamic equations to the extent that they become sufficiently simple for an analytical treatment, and secondly to define the range of applicabilty of these reduced equations.

Why speaking of gases, not of fluids in general? The reason is that the leading equations in hydrodynamics, the Euler and Navier--Stokes equations, are valid for ideal gases only. This statement is likely to be met with scepticism, since it appears to be general understanding that these equations cover the full range from ideal gases via real gases to liquids, differing only with regard to their values of the transport coefficients. One has to realize, however, that the system of hydrodynamic transport equations is based on the kinetic equation for the distribution function in 6--dimensional phase space (see, e.g.~[1],[2],[3]). It is a built--in assumption in the kinetic equation that the particles are points without finite volume, so that an infinite number of particles can be accomodated in a finite volume. Obviously, this assumption fails badly for liquids. Furthermore, in order to arrive at the Euler or Navier--Stokes equations in their known forms, it is a necessity to disregard intermolecular forces in the kinetic equation. 

The kinetic equation, as well as everything that follows from it, will have to be modified considerably if the finite volume of  particles and the intermolecular forces between them are taken into account. A certain insight may be obtained by using van der Waals' equation of state to reformulate the Euler equation, leading to  

\begin{equation} 
\frac{\partial \bf v}{\partial t}  + ({\bf v} \cdot \nabla ){\bf v} =  - \, V^2 \, \varphi(N) \left[\, \frac{\nabla T}{T} + \psi(N,T) \frac{\nabla N}{N} \, \right] 
\end{equation}

where $V$ is the thermal velocity ($V^2 = KT/m$, with $K$ the Boltzmann constant) and $N$ is the number density. The functions $\varphi(N)$ and $\psi(N,T)$ have the properties $\varphi$, $\psi \rightarrow 1$ when the ideal gas state is approached, and  $\varphi$, $ \psi \rightarrow \infty$ when the state of densest possible package of particles in a liquid is reached. Eq.~(1) shows that the density gradient term has a higher weight as a driving agent for the velocity field than the temperature gradient term, except for an ideal gas.  

Eq.~(1) gives an indication of how important it would be to find a formulation of the kinetic equation, and of the ensuing transport equations, that would incorporate both the finite size of particles and the intermolecular forces between them. Here we will have to be satisfied with speaking about ideal gases.

\vspace*{4mm} {\large \bf 2. Euler and Navier--Stokes systems}

For easier reference in the later part of this paper, we write down the equations belonging to the Euler and to the Navier--Stokes system. It is important to realize that these systems consist of three equations each, for the zero--order moment $N$, the first--order moment {\bf v}, and the second--order moment $T$. This three--equation system is backed by joint truncation prescriptions, needed to convert the infinite system of transport equations into a usable finite set (see [1],[2],[3]). 

The Euler system is given by (e.g., see eqs.~(24) and (25) of [3])

\begin{equation} 
\frac{\partial N}{\partial t} + {\bf v} \cdot \nabla N  = -\, N \, \nabla \cdot {\bf v}
\end{equation}
\begin{equation} 
\frac{\partial \bf v}{\partial t}  + ({\bf v} \cdot \nabla ){\bf v} =  - \, V^2 \left[\, \frac{\nabla T}{T} + \frac{\nabla N}{N} \, \right] 
\end{equation}
\begin{equation} 
\frac{\partial T}{\partial t} + {\bf v} \cdot \nabla T = - \frac{2}{3} \, T \, \nabla \cdot {\bf v}
\end{equation}

Since these equations relate to an ideal gas, the pressure $p(N,T)$ is given by 

\begin{equation} 
p = NKT
\end{equation}

and hence the RHS term in (3) can be written as $ - 1 / (Nm) \nabla p $, whereby (3) would adopt the conventional form of the Euler equation.

The Navier--Stokes system is given by eq.~(2) and (see eqs.~(41) and (43) of [3]):

\begin{equation} 
\frac{\partial \bf v}{\partial t}  + ({\bf v} \cdot \nabla ){\bf v} =   - \,  V^2 \left[\, \frac{\nabla T}{T} + \frac{\nabla N}{N} \, \right] +\frac{\eta}{N m} \, \nabla \cdot  \left[\, \nabla {\bf v} + (\nabla {\bf v})^t - \frac{2}{3}\, (\nabla\cdot {\bf v}) \, \textsf{U} \, \right]
\end{equation}
\begin{equation} 
\frac{\partial T}{\partial t} + {\bf v} \cdot \nabla T = -\frac{2}{3}\, T \, \nabla \cdot {\bf v}
+ \frac{2}{3}\, \frac{\eta}{KN}\, \left[ \nabla {\bf v} : [\nabla {\bf v} + (\nabla {\bf v})^t] - \frac{2}{3} (\nabla\cdot {\bf v})^2 \right] + \frac{2}{3}\, \frac{\kappa}{KN}\, \nabla \cdot (\nabla T)
\end{equation}

Here $(\nabla {\bf v})^t$ denotes the transposed dyadic, $\sf U$ is the unit tensor, $\eta$ the dynamic viscosity and $\kappa$ the heat conductivity.\footnote{The thermal equation in its original form involves the 27--element heat flow tensor. However, by means of simplifying assumptions corresponding to those leading to the Navier--Stokes equation (6), the effect of the heat flow tensor can be reduced to the term $\kappa \, \nabla \cdot (\nabla T)$ in (7). In this way, the infinite set of transport equations is truncated behind second order moments and thereby reduced to the closed three--equation system [(2),(6),(7)]; see [3].}

Both systems rely on the strong isotropizing action of collisions. In the case of the Euler system [(2),(3),(4)] the condition is that the 9--element pressure tensor $\sf p$ can be reduced to ${\sf p} = p\, {\sf U}$. In the case of the Navier--Stokes system [(2),(6),(7)] this condition is relaxed to $| p_{ij}| \ll p_{ii} $ and $| p_{ii} - p| \ll p$. These conditions invalidate solutions far away from equilibrium as, for instance, so--called blowup solutions. Another condition is that the mean free path is the shortest length in the system (small Knudsen number), and the time between collisions the shortest time\footnote{The complete applicability condition of fluid theory reads $\tau^{-2}+\tau_c^{-2}\gg V^2/l^2$, where $\tau$ is a characteristic time and $l$ a characteristic length of the given process, and $\tau_c$ is the average time between collisions; see [4]. It follows that $\tau_c \ll \tau$ is the precondition for the Knudsen condition $\tau_c V \ll l$ (collisional localization). Fluid theory is also applicable for extremely rapid processes, satisfying $\tau V \ll l$ (inertial localization).}. Furthermore it is as\-sumed in the derivation that the particles have no internal degrees of freedom. Otherwise separate kinetic equations would have to be formulated for every internal excitation level, with transition terms between them. However, in the final results above the factor 2/3 may be translated into $\gamma - 1$, with $ \gamma$ the actual value of the adiabatic cofficient, subject to the number of degrees of freedom, provided these are fully activated.

Both systems are complete. No additional equation is needed, and if attempts are made to introduce an additionl equation, no empty place is found. If a place is cleared by removing one of the existing equations -- eqs.~(4) or (7) are typical candidates -- the path of valid physics is left. This remark is of specific relevance when the incompressibility condition $\nabla \cdot {\bf v} = 0$ is used in  place of either (4) or (7). The physical consequences of this procedure will be discussed in Sections 4 and 5 below.

If the above transport equations are simplified by employing the property of incompressibility, it will be necessary to have criteria  allowing to check the physical validity of the resulting simplified transport equations. An absolute criterion is given by the requirement that the total energy inside a closed solid surface is conserved. Further criteria can be obtained by considering the various occurring energy transfer rates.   

For this purpose we convert (6) and (7) into equations for the kinetic energy density $w_K = \frac{1}{2} Nm{\bf v}^2$  and the internal energy density $w_I = \frac{3}{2} NKT = \frac{3}{2}\, p$, and we obtain (see eqs.~(49)--(51) of [3]):

\begin{equation} 
\frac{\partial w_K}{\partial t} + \nabla \cdot ( w_K \, {\bf v}) = - \, {\bf v}\cdot \nabla p +  \eta \, {\bf v}\cdot  \left( \nabla \cdot \left[\, \nabla {\bf v} + (\nabla {\bf v})^t - \frac{2}{3}\, (\nabla\cdot {\bf v}) \, \textsf{U} \, \right] \right)
\end{equation}
\begin{equation} 
\frac{\partial w_I}{\partial t} + \nabla \cdot (w_I \, {\bf v}) = - \,   p \, \nabla \cdot {\bf v} + \eta \, \left[ \nabla {\bf v} : [\nabla {\bf v} + (\nabla {\bf v})^t] - \frac{2}{3} (\nabla\cdot {\bf v})^2 \right] + \kappa \, \nabla\cdot (\nabla T) 
\end{equation}
\begin{equation} 
\frac{\partial w}{\partial t} + \nabla \cdot (w \, {\bf v}) = - \nabla \cdot (p \, {\bf v}) +  \eta \, \nabla \cdot  \left( {\bf v} \cdot \left[\, \nabla {\bf v} + (\nabla {\bf v})^t - \frac{2}{3}\, (\nabla\cdot {\bf v}) \, \textsf{U} \, \right] \right) + \kappa \, \nabla\cdot (\nabla T) 
\end{equation}

where $w = w_K + w_I$. With Gau{\ss}' integral theorem, a consequence of (10) is that the total energy $ \int w\, d^3{\bf r}$ 
\hspace*{1mm}inside a closed solid surface, or in entire space, is constant, 

\begin{equation} 
\frac{d}{dt}\int w \, d^3 {\bf r} = 0
\end{equation}

and this is the minimal condition that any simplified version of the transport equations has to satisfy.

The energy equations (8) and (9) can be written in the transparent form (see eqs.~(28) and (29) of [5])

\begin{equation} 
\frac{\partial w_K}{\partial t} + \nabla \cdot ( w_K
 \, {\bf v})  =  \dot{w}_{K\leftrightarrow P} + \dot{w}_{K\leftrightarrow K} - \dot{w}_{K\rightarrow I}
\end{equation}
\begin{equation} 
\frac{\partial w_I}{\partial t} + \nabla \cdot (w_I \, {\bf v}) =
\dot{w}_{I\leftrightarrow P} + \dot{w}_{I\leftrightarrow I} +
\dot{w}_{K\rightarrow I}
\end{equation}

The energy transfer rates on the right--hand sides of (12) and (13) are defined by eqs.~(23)--(27) of [5], and they have the following physical meanings: $\dot{w}_{K\leftrightarrow P}$ describes the mutual conversion of kinetic and potential energy of an ensemble of particles moving along or opposite to the pressure gradient, and $\dot{w}_{I\leftrightarrow P}$ describes the mutual conversion of internal and potential energy due to compression or expansion. The terms $\dot{w}_{K\leftrightarrow K}$ and $\dot{w}_{I\leftrightarrow I}$ express the spatial redistribution of kinetic and internal energy under the action of viscosity and heat conduction, respectively, and $\dot{w}_{K\rightarrow I}$ describes the irreversible conversion of kinetic into internal energy.

Insertion of the Navier--Stokes approximation for the pressure tensor and the heat flux vector (see eqs.~(37) and (39) of [3]) into the defining equations for the energy transfer rates (eqs.~(23)--(27) of [5]) yields

\begin{equation} 
\dot{w}_{K\leftrightarrow P} = - \, {\bf v} \cdot \nabla p + 2 \, \eta \left[ v_x \frac{\partial^2 v_x}{\partial x^2} +  v_y \frac{\partial^2 v_y}{\partial y^2} +  v_z \frac{\partial^2 v_z}{\partial z^2} \, \right] - \frac{2}{3} \eta \, \, {\bf v} \cdot \nabla ( \nabla \cdot {\bf v})
\end{equation}
\begin{equation} 
\dot{w}_{I\leftrightarrow P} = - \, p \, \nabla \cdot {\bf v} + 2 \, \eta \bigg[ \left( \frac{\partial v_x}{\partial x} \right)^{\mbox{\hspace*{-1.2mm}}2} + \left( \frac{\partial v_y}{\partial y} \right)^{\mbox{\hspace*{-1.2mm}}2} + \left( \frac{\partial v_z}{\partial z} \right)^{\mbox{\hspace*{-1.2mm}}2} \, \bigg] - \frac{2}{3} \eta \, \, (\nabla \cdot {\bf v} )^2
\end{equation}
\begin{equation} 
\dot{w}_{K\leftrightarrow P} + \dot{w}_{I\leftrightarrow P} = - \, \nabla \cdot (p \, {\bf v}) + 2 \, \eta \left[ \frac{\partial}{\partial x} \left( v_x \frac{\partial v_x}{\partial x} \right) + ..... + ..... \,\right] - \frac{2}{3} \eta \, \nabla \cdot [{\bf v} \, ( \nabla \cdot {\bf v}) ]
\end{equation}
\begin{equation} 
\dot{w}_{K\leftrightarrow K} = \eta \left\{ \frac{\partial}{\partial x} \left[ v_y \left( \frac{\partial v_x}{\partial y} + \frac{\partial v_y}{\partial x} \right) + v_z \left( \frac{\partial v_x}{\partial z} + \frac{\partial v_z}{\partial x} \right) \right] 
+ ..... + ..... \, \, \right\}
\end{equation}
\vspace*{-1mm}
\begin{equation} 
\dot{w}_{I\leftrightarrow I} = \kappa \, \nabla\cdot (\nabla T) 
\end{equation}
\vspace*{-1mm}
\begin{equation} 
\dot{w}_{K\rightarrow I} = \eta \, \bigg[ \left( \frac{\partial v_x}{ \partial y} 
+ \frac{\partial v_y}{\partial x} \right)^{\mbox{\hspace*{-1.2mm}}2} + \left( \frac{\partial v_y}{\partial z} + \frac{\partial v_z}{\partial y} \right)^{\mbox{\hspace*{-1.2mm}}2} +\left( \frac{\partial v_z}{\partial x} + \frac{\partial v_x}{\partial z} \right)^{\mbox{\hspace*{-1.2mm}}2} \, \bigg]
\end{equation}

We conclude from (16)--(18):

\begin{equation} 
\int (\dot{w}_{K\leftrightarrow P} + \dot{w}_{I\leftrightarrow P}) \, d^3 {\bf r} = 0
\end{equation}
\begin{equation} 
\int \dot{w}_{K\leftrightarrow K} \, d^3 {\bf r} = 0 
\end{equation}
\begin{equation} 
\int \dot{w}_{I\leftrightarrow I} \, d^3 {\bf r} = 0 
\end{equation}

Furthermore it follows from (19) that $ \dot{w}_{K\rightarrow I} $ is an unconditionally positive quantity, thereby expressing the irreversible nature of the conversion of kinetic into internal energy. 

The relations (11) and (20)--(22) constitute conditions that will have to be satisfied by any simplified system of transport equations.   

\vspace*{4mm} {\large \bf 3. Meaning of incompressibility}

There is no need to consider the incompressible case separately, because incompressible behaviour, if existing, will follow self--consistently from the full equations. The motivation stems from the wish to obtain simple equations, in order not to depend on numerical methods. On this way, slight simplifications will not be helpful, one has to obtain equations simple enough to be amenable to an analytical treatment, which may conflict with the need not to violate physical essentials. 

If one understands incompressibily literally as the inability of a substance to reduce its volume as a reaction upon squeezing, then this property of the substance forces the velocity field into the property $\nabla \cdot {\bf v} = 0$. But what if the given substance, an ideal gas, is far from being incompressible in its literal sense? In this case one would have to turn the argument around by saying that the condition $\nabla \cdot {\bf v} = 0$ has to be imposed on the equations in order to enforce solutions which make the substance appear as if it were incompressible. 

The simplification achieved by $\nabla \cdot {\bf v} = 0$ is not sufficient in a practical sense since the coupling of the equations (3) and (6) with the continuity equation remains. Additionally it has to be demanded that $ \nabla N = 0 $. The incompressibility condition is thus given by

\begin{equation} 
\nabla \cdot {\bf v} = 0  \mbox{ \hspace*{5mm} {\rm and} \hspace*{5mm}}   \nabla N = 0 
\end{equation}

with the consequence that $N$ has to be constant in space and time. The equation of state thereby becomes 

\begin{equation} 
p = N_0KT 
\end{equation}

with $N_0$ the constant value of $N$.

By order of magnitude estimates  it has been demonstrated (e.g., \S 10 of [6]) that the condition (23), applied to the Euler equation, has a good physical background if both the flow velocity and the propagation velocity of a perturbation are much smaller than the sound velocity $V_s$,

\begin{equation} 
v \ll V_s \mbox{ \hspace*{5mm} {\rm and} \hspace*{5mm}}   \frac{l}{\tau} \ll V_s  
\end{equation}

where $l$ is a length and $\tau$ a time by which a given process can be characterized. However, since these arguments are based on the Euler equation alone, they are not applicable to the Navier--Stokes equation and to the thermal equation. Moreover, they do not include a prescription for how to incorporate the incompressibility condition (23) in the hydrodynamic equations. There exist different views on how to do this.

\vspace*{4mm} {\large \bf 4. Euler system with incompressibility}

Application of the incompressibility condition (23) to the equations (3) and (4) yields

\begin{equation} 
\frac{\partial \bf v}{\partial t}  + ({\bf v} \cdot \nabla ){\bf v} =  - \, V^2 \, \frac{\nabla T}{T}  
\end{equation}
\begin{equation} 
\frac{\partial T}{\partial t} + {\bf v} \cdot \nabla T = 0
\end{equation}

This is again a closed set of equations, describing the two remaining variables {\bf v} and $T$, and since the incompressibility condition (23) is fully incorporated in (26) and (27), one should think that nothing could be left to be said.

In fact, however, the equations (26) and (27) are not meant when the so--called ``incompressible Euler equations'' are addressed. These  are given by 

\begin{equation} 
\frac{\partial \bf v}{\partial t}  + ({\bf v} \cdot \nabla ){\bf v} =  - \frac{1}{N_0m} \nabla p  
\end{equation}
in conjunction with 
\begin{equation} 
\nabla \cdot {\bf v} = 0  
\end{equation}

The system [(28),(29)] is commonly used in the mathematically oriented literature (see the review by Constantin, [7]). By virtue of (24) it is seen that eqs.~(26) and (28) are equivalent.

We now have two systems of equations, [(26),(27)] and [(28),(29)], serving the same purpose, but differing with respect to their closure equations, (27) and (29) respectively. Which of the two systems should be preferred? The answer is none, both have serious defects:  

The physical meaning of (27) is that in the absence of compression or expansion the only way to change the local value of $T$ is by  means of convection. Solutions of [(26),(27)] will meet this requirement, but they will not necessarily show incompressible behaviour. The use of the incompressibility assumption in establishing (26) and (27) is not sufficient to secure that the solutions will actually confirm the validity of this assumption. This possible contradiction between the end result and the starting assumption can be resolved only by selecting those solutions which satisfy both (27) and (29). 

Corresponding arguments apply to (29): Using (29), the solutions of [(28),(29)] will show incompressibility, but they will not necessarily satisfy (27). So we see that the enforcement of incompressibility by means of (29) will lead to physically valid results only if (27) is satisfied at the same time. 

An iterative procedure will be one possibility to attempt finding solutions which simultaneously satisfy (26) or (28) together with (27) and (29), which means to have three equations for the two variables {\bf v} and $T$ or $p$. However, why should one make the assumption of incompressibility, aiming at simplifying matters, when in the end such an entirely impractical method should be necessary.

A second way to reconcile the two closure relations (27) and (29) is to find a condition for that both are not needed, and this is by disregarding the pressure gradient term on the RHS of (28) in comparison with the LHS terms, whereby the Euler equation is converted into the cold fluid equation  $ \partial {\bf v}/\partial t + ({\bf v} \cdot \nabla ){\bf v} = 0 $. This equation appears rather useless, but some significance can be given to it by introducing an external force {\bf F},  

\begin{equation} 
\frac{\partial \bf v}{\partial t}  + ({\bf v} \cdot \nabla ){\bf v} =  \frac{\bf F}{m}
\end{equation}

The force {\bf F} may contain an externally impressed pressure gradient in which case (30) would look as if it were the Euler equation. The obvious shortcoming of (30) is that it disregards the coupling with other fluid variables. On the other hand, (30) has the advantage that its applicability is not restricted to ideal gases, and this may explain its success in certain practical applications.

The term `cold fluid' has been adopted here from plasma physics where it is used when the effect of Coulomb and Lorentz forces dominates over kinetic effects. The cold fluid equations, of which (30) is a part, form an infinite set which can be obtained from the kinetic equation. Following the treatment in Section 2 of [3], the term in the kinetic equation which is the source of all the complexities in the system of transport equations is $({\bf u} - {\bf v})\cdot \nabla \tilde{f}$ (with $\tilde{f}$ the distribution function in velocity space), and the cold fluid approximation is obtained if this term is neglected. Thereby, the following set of cold fluid equations is obtained: 

\begin{equation}\tag{31.1} 
\frac{\partial \bf v}{\partial t}  + ({\bf v} \cdot \nabla ){\bf v} =  \frac{\bf F}{m} 
\end{equation}
\begin{equation}\tag{31.2}
\frac{\partial T}{\partial t} + {\bf v} \cdot \nabla T = 0
\end{equation}
\begin{equation}\tag{31.3}
\frac{\partial p_{ii} }{\partial t} + {\bf v} \cdot \nabla p_{ii}  - \frac{\delta (p_{ii} - p)}{\delta t}  = 0
\end{equation}
\begin{equation}\tag{31.4}
\frac{\partial p_{ij}}{\partial t} + {\bf v} \cdot \nabla p_{ij} - \frac{\delta p_{ij}} {\delta t}  = 0
\end{equation}
 
and so on. Here $\delta / \delta t$ denotes the temporal change due to the randomizing action of self-collisions. Eq.~(31.2) is identical with (27), i.e., application of $\nabla \cdot {\bf v} = 0$ in the thermal equation (4) corresponds to using the cold fluid equation (31.2). It should be added that the continuity equation (2), being an exact equation outside the reach of simplifications in the kinetic equation, remains unchanged. The cold fluid equations above hold no possibility to incorporate the condition of incompressibility.

In concluding this section, the ``incompressible Euler equations'' [(28),(29)] are physically valid only in the accidental case that their solutions satisfy (27), too, and hence they are of no practical use, notwithstanding their potential role as an interesting mathematical research object (e.g., see [7]). Likewise, the equations [(26),(27)] are physically valid only in the accidental case that their solutions satisfy (29), too. As a consequence, incompressibility is not a property that can be enforced. There exists no alternative to using the full Euler system [(2),(3),(4)] or, if justifiable in a given physical situation, the cold fluid equations. 

\setcounter{equation}{31}

\vspace*{4mm} {\large \bf 5. Navier--Stokes system with incompressibility}

After we have seen that the assumption of incompressibility is of no use to simplify the Euler system in a justified way, it would be unrealistic to expect that the situation should be better in the case of the Navier--Stokes system.

Application of the incompressibility condition (23) to the Navier--Stokes equations (6) and (7) yields 

\begin{equation} 
\frac{\partial \bf v}{\partial t}  + ({\bf v} \cdot \nabla ){\bf v} =   - \,  \frac{K}{m} \, \nabla T +\frac{\eta}{N_0 m} \, \nabla \cdot   [ \nabla {\bf v} + (\nabla {\bf v})^t]
\end{equation}
\begin{equation} 
\frac{\partial T}{\partial t} + {\bf v} \cdot \nabla T =  \frac{2}{3}\, \frac{\eta}{KN_0}\,  \nabla {\bf v} : [\nabla {\bf v} + (\nabla {\bf v})^t] + \frac{2}{3}\, \frac{\kappa}{KN_0}\, \nabla \cdot (\nabla T)
\end{equation}

The energy exchange rates $\dot{w}_{K\leftrightarrow K}$, $\dot{w}_{I\leftrightarrow I}$ and $\dot{w}_{K\rightarrow I}$ given by (17)--(19) remain unchanged, whereas $\dot{w}_{K\leftrightarrow P}$ and $\dot{w}_{I\leftrightarrow P}$ are changed to

\begin{equation} 
\dot{w}_{K\leftrightarrow P} = - \, \nabla \cdot (p \, {\bf v}) + 2 \, \eta \left[ v_x \frac{\partial^2 v_x}{\partial x^2} + v_y \frac{\partial^2 v_y}{\partial y^2} + v_z \frac{\partial^2 v_z}{\partial z^2} \, \right]
\end{equation}
\begin{equation} 
\dot{w}_{I\leftrightarrow P} =  2 \, \eta \, \bigg[ 
\left( \frac{\partial v_x}{\partial x} \right)^{\mbox{\hspace*{-1.2mm}}2}  
+ \left( \frac{\partial v_y}{\partial y} \right)^{\mbox{\hspace*{-1.2mm}}2}  
+ \left( \frac{\partial v_z}{\partial z} \right)^{\mbox{\hspace*{-1.2mm}}2}  
\, \bigg] 
\end{equation}
\begin{equation} 
\dot{w}_{K\leftrightarrow P} + \dot{w}_{I\leftrightarrow P} = - \, \nabla \cdot (p \, {\bf v}) + 2 \, \eta \left[ \frac{\partial}{\partial x} \left( v_x \frac{\partial v_x}{\partial x} \right) + \frac{\partial}{\partial y} \left( v_y \frac{\partial v_y}{\partial y} \right) + \frac{\partial}{\partial z} \left( v_z \frac{\partial v_z}{\partial z} \right) \,\right]
\end{equation}

The relation (35) is problematic since it implies that $\dot{w}_{I\leftrightarrow P}$ is positive always and everywhere. This corresponds to a one--sided conversion of potential into internal energy by compression, where actually the conversion should be mutual and comprise both compressions and expansions. This problem can be remedied by narrowing the incompressibility condition (23) to the non--deformability condition

\begin{equation} 
\frac{\partial v_x}{\partial x} = \frac{\partial v_y}{\partial y} = \frac{\partial v_z}{\partial z} = 0 \mbox{ \hspace*{5mm} {\rm and} \hspace*{5mm}}   \nabla N = 0 
\end{equation}

so that only changes of {\bf v} perpendicular to its own direction would be allowed.

As a consequence of (37) we have $\dot{w}_{K\leftrightarrow P} = - \, \nabla \cdot (p \, {\bf v})$ and $\dot{w}_{I\leftrightarrow P} = 0$. A further consequence is $p_{xx} = p_{yy} = p_{zz} = p$, and the dyadic $\nabla {\bf v}$ is reduced to $(\nabla {\bf v})_0$ which originates from $\nabla {\bf v}$ by setting the elements in the diagonal to zero. With (37) in place of (23), eqs.~(32) and (33) are thus altered to  

\begin{equation} 
\frac{\partial \bf v}{\partial t}  + ({\bf v} \cdot \nabla ){\bf v} =   - \, \frac{K}{m} \nabla T +\frac{\eta}{N_0 m} \, \nabla \cdot  [\, (\nabla {\bf v})_0 + (\nabla {\bf v})_0^t]
\end{equation}
\begin{equation} 
\frac{\partial T}{\partial t} + {\bf v} \cdot \nabla T =  \frac{2}{3}\, \frac{\eta}{KN_0}\, \, (\nabla {\bf v})_0 : [\, (\nabla {\bf v})_0 + (\nabla {\bf v})_0^t \, ] + \frac{2}{3}\, \frac{\kappa}{KN_0}\, \nabla \cdot (\nabla T)
\end{equation}

This appears to be the simplest possible form of the Navier--Stokes system, but it is still not simple enough to be useful in a practical sense, hardly easier to handle than the full Navier--Stokes system [(2),(6),(7)]. It appears pointless, therefore, to sacrifice general validity for the minor simplifications achieved thereby.

Actually, an even simpler version is used in parts of the literature, given by the so--called ``incompressible Navier--Stokes equations'', consisting of 

\begin{equation} 
\frac{\partial \bf v}{\partial t}  + ({\bf v} \cdot \nabla ){\bf v} =   - \, \frac{1}{N_0m} \nabla p +\frac{\eta}{N_0 m}  \Delta {\bf v} 
\end{equation}
and
\setcounter{equation}{28}
\begin{equation} 
\nabla \cdot {\bf v} = 0
\end{equation}
\setcounter{equation}{40} 

\vspace*{-4mm} Eq.~(40) follows from (32) with (24) and the identity $\nabla \cdot  [\, \nabla {\bf v} + (\nabla {\bf v})^t] = \Delta {\bf v} + \nabla ( \nabla \cdot {\bf v} )$. The system [(40),(29)] is frequently used in the mathematically oriented literature, and the Millennium Prize endowed by the Clay Mathematics Institute (Fefferman [8]) is a prominent example. 

The problem here is the same as in the case of the Euler system: The thermal equation is removed and replaced by (29), yet the thermal equation is an irremovable part of the Navier--Stokes system which does not lose its existence only because incompessibility is assumed (see [3]). Consequently, (29) can be used only as an additional equation, whereupon there would be three equations for two unknowns. The aim to achieve simplification by assuming incompressibility would thus be missed and changed to the opposite. Incompressibility is a subordinate property, potentially following from the full set of transport equations, but unable to dictate the theoretical description from the beginning.

Additionally, the system [(40),(29)] has the following specific physical deficiencies: 

1. The absence of a thermal equation has the consequence that possible local accumutations or depletions of internal energy cannot be smoothed out by means of heat conduction ($ \dot{w}_{I\leftrightarrow I} = 0 $).

2. The system [(40),(29)] contains no term to describe the mutual conversion of potential and internal energy ($ \dot{w}_{I\leftrightarrow P} = 0 $), with the consequence that condition (20) is not obeyed, except for the case that the non--deformability condition (37) is satisfied. 

3. The loss of kinetic energy by irreversible conversion into internal energy, described by the term $ \dot{w}_{K\rightarrow I} $ in the energy equation (12), is not balanced anywhere in the system, with the consequence that the total energy is not conserved, 

\begin{equation} 
\frac{d}{dt} \int w \, d^3 {\bf r} \,\, < \,\, 0
\end{equation}

This non--conservation of energy is an absolute disqualifier for the system [(40),(29)].

It is astonishing how uncritically the incompressibility condition $\nabla \cdot {\bf v} = 0$ is used in parts of the literature, particularly so if this condition is used in place of the thermal equation. We have seen that incompressibility is not a property that can be enforced in a physically justified way by means of an external condition. Incompressible be\-ha\-viour, if existing, will have to follow self--consistently from the full system of transport equations. It may be regrettable that these equations are too complex for an analytical treatment, but the wish for mathematical simplicity is not a sufficient argument for mutilating physical equations as, for instance, in the case of the Millennium Prize problem.

\vspace*{2cm} 
\centerline {\bf References}
\vspace*{3mm}

[1] S.~Chapman and T.~Cowling, {\it The Mathematical Theory of Non--Uniform
Gases}, Cambridge University Press (1970).

[2] R.~Schunk, {\it Mathematical structures of transport equations for multispecies flow}, Rev.~Geophys.~Space Phys.~{\bf 15} (1977), 429--445. 

[3] P.~Stubbe, {\it The Euler and Navier--Stokes equations revisited}, arXiv:1506.04561 (2015). 

[4] P.~Stubbe. {\it The concept of a kinetic transport theory}, Phys.~Fluids B {\bf 2}(1) (1990), 22--33.

[5] P.~Stubbe, {\it Note on Onsager's conjecture}, arXiv:1509.08406 (2015). 

[6] L.~Landau and E.~Lifshitz, {\it Course of Theoretical Physics, Vol.~6: Fluid Mechanics}, Pergamon Press (1987).

[7] P.~Constantin, {\it On the Euler equation of incompressible fluids}, Bull.~Amer.~Math.~Soc.~{\bf 44} (2007), 603--621.

[8] C.~Fefferman, {\it Existence and smoothness of the Navier--Stokes equation}, http://www.claymath.org/
millennium/Navier-Stokes\_Equations/navierstokes.pdf

\end{document}